\documentclass[english]{article}
\usepackage[T1]{fontenc}
\usepackage[latin9]{inputenc}
\usepackage{geometry}
\usepackage{amsmath}
\usepackage{amssymb}
\usepackage{graphicx}
\usepackage[all,cmtip]{xy}

\makeatletter
\date{}

\makeatother

\usepackage{babel}
\begin{document}
\title{\textbf{Canonical Group Quantization of Noncommutative Graphene with Symmetric and Landau Dual Magnetic Fields}}
\author{\vspace{5mm}
\textbf{M. F. Umar}\textsf{}\thanks{\textsf{faudzi@fsmt.upsi.edu.my}} $^{a}$ and
\textbf{M. S. Nurisya}\thanks{\textsf{risya@upm.edu.my}} $^{b,c}$}
%and \textbf{H. Zainuddin}$^{a,b}$}
\maketitle
\begin{center}
{$^{a}$\em Department of Physics, Faculty of Science and Mathematics,  }\\
{\em Universiti Pendidikan Sultan Idris, }\\
{\em 435900 Tanjong Malim, Perak, Malaysia.}
\par\end{center}
\begin{center}
{$^{b}$\em Laboratory of Computational Sciences and Mathematical
Physics, }\\
{\em Institute for Mathematical Research (INSPEM), Universiti Putra
Malaysia,}\\
{\em 43400 UPM Serdang, Selangor, Malaysia.}
\par\end{center}

\begin{center}
{$^{c}$\em Department of Physics, Faculty of Science, Universiti
Putra Malaysia, }\\
 {\em 43400 UPM Serdang, Selangor, Malaysia.}
\par\end{center}

\begin{center}
\par\end{center}

\begin{center}
\vspace{0.5mm}
\par\end{center}
\begin{abstract}
The canonical group quantization approach has been used to study noncommutative graphene in the presence of dual magnetic fields. The canonical group for the phase space $\mathbb{R}^2\times \mathbb{R}^2$ with both symmetric and Landau dual gauges is shown to be equivalent to $\mathtt{H}^2\rtimes \mathbb{R}$. The representations of both symmetric and Landau dual gauges lead to similar canonical commutation relations, and we observe that the energy spectrum is corrected by both dual magnetic fields, yielding the same result.

\end{abstract}

\section{Introduction}

Graphene, a 2D configuration of carbon atoms, has captured significant attention in various branches of physics \cite{novoselov2004electric,zhang2005experimental,peres2010colloquium}. The behavior of electrons in graphene is similar to that of massless Dirac fermions, exhibiting properties of relativistic particles. The fact that electrons in graphene display relativistic behavior creates opportunities for exploring relativistic phenomena in condensed matter. The most fascinating phenomena in graphene are the anomalous Landau-Hall effect \cite{meyer2007structure} and Klein-paradox \cite{neto2009electronic}. In other words, graphene serves as a link between condensed matter and high-energy physics.

The study of quantum systems in a noncommutative space has been the subject of much interest at very tiny scales and very high energies \cite{seiberg1999string}. Some attention has been given to the models of noncommutative quantum mechanics (NCQM). NCQM is an attractive research field because it allows physicists to better understand the implications of noncommutativity by using conventional quantum mechanics calculations, and can be achieved by several formulations \cite{gouba2016comparative} \textit{i.e.} Bopp shift, Moyal $\star$-product and Seiberg Witten map. NCQM has explored many physical systems such as harmonic oscillator \cite{li2005representation}, quantum Hall effect \cite{horvathy2006anomalous}, and Landau effect \cite{gangopadhyay2015landau}.

Contravariant gauge is the contravariant geometric on gauge, and been used in emergent gravity \cite{kaneko2018contravariant}. The study on contravariant gauge (or dual gauge) also known as noncommutative gauge, has appeared to be very encouraging, particularly in terms of the group structure.  In literature, the author includes the minimal coupling of positions and momenta with dual magnetic and magnetic field respectively \cite{ngendakumana2014group}. However, our work  considers both symmetric and Landau dual gauges to explore the consequences of the dual magnetic fields into quantum systems. 

This study uses the method of quantization proposed by C. J. Isham \cite{isham1984topological,isham1984groupa,isham1984groupb}, known as group-theoretic or canonical group quantization (CGQ), This approach involves a geometric approach that incorporates the group structure as a central element in the process. The first step involves identifying the canonical group that describes the symmetries of the  phase space. The generators of the group that correspond to classical observables are then quantized by determining their irreducible unitary representations. By applying this method to the linear phase space $\mathbb{R}^2$, the well-known canonical commutation relation (CCR) is obtained, and conventional quantum mechanics can also be obtained
\begin{subequations}
 \begin{align}\label{1a}
[\hat{q}^i, \hat{q}^j]&=0, \\ [\hat{p}_i, \hat{p}_j]&=0,\\ [\hat{q}^i, \hat{p}_i]&=i\hbar,
\end{align}   
\end{subequations}     where $i, j = 1, 2$ for a two-dimensional plane as the configuration space and $\hbar$ is the reduced Planck constant. Noncommutative positions will change (\ref{1a}) that has been studied by \cite{umar2dpmssaq} using CGQ. However, the symmetric and Landau dual magnetic fields in NCQM will be delved into in the first segment using this quantization approach on noncommutative plane.
 
Some authors have investigated the behavior of graphene in a noncommutative space, such as exploring the Landau levels of both single-layer and bilayer graphene within this framework \cite{sourrouille2021landau}. In a subsequent study, the noncommutative phase-space extension of graphene under the influence of an external, constant magnetic field was analyzed \cite{bastos2013noncommutative}. The researchers found that the noncommutativity of momenta led to corrections in the energy spectrum of the system. However, the introduction of a dual magnetic background to graphene may present a new and exciting research opportunity. This motivation leads us to identify the canonical group for the phase space with symmetric and Landau dual magnetic fields. To accomplish the quantization of noncommutative graphene, we proceed to find its representations. We then compute the energy spectrum of the noncommutative graphene and compare the effect of both dual magnetic fields. 

The present article is organized as follows. In Sec. \ref{cgqancp}, we give the basic outline of CGQ to develop NCQM with dual magnetic field background. In Sec. \ref{ncg}, noncommutative graphene is studied using the representations from the previous section. We compute the energy spectrum of the noncommutative graphene, and brifely discuss the comparison between dual magnetic fields. Finally we conclude our findings in Sec. \ref{4aaa}.

\section{Canonical Group Quantization and Noncommutative Plane $\mathbb{R}^2$}\label{cgqancp}

Canonical group quantization (CGQ) or Isham's quantization method briefly has two steps namely canonical group and irreducible unitary representations. Canonical group $\mathcal{C}$ is a Lie group $\mathcal{G}$ that describes the symmetries of the phase space of the system $\mathcal{S}$, and the quantization ends the procedure by seeking all inequivalent irreducible unitary representations of $\mathcal{C}$ that correspond to the classical observables that are to be quantized\footnote{Note that $\mathcal{C}$ denoted as the canonical group and $\mathcal{G}$ for Lie group}. 

For the canonical group stage, it starts with a canonical Lie algebra $\mathcal{L}(\mathcal{C})$ merely is a Lie algebra $\mathcal{L}(\mathcal{G})$ of the phase space, in which each element will induce  vector field $\gamma$ via the group action on the phase space, while under moment map $P$, the element develops the classical observable of the phase space $C^{\infty}(\mathcal{S},\mathbb{R})$. Classically, both  (Hamiltonian) vector field of the phase space $HamVF(\mathcal{S})$ and classical observables are linked with one another via $j$ map. 

\begin{align}\label{2}
\xymatrix{
	C^{\infty}(\mathcal{S},\mathbb{R})\ar[r]^j &HamVF(\mathcal{S})
	\\\mathcal{L}(\mathcal{C})\ar@{.>}[u]^P &\mathcal{L}(\mathcal{G})\ar@{~>}[l]^{\simeq}\ar[u]^{\gamma}
\\\mathcal{C}\ar@{.>}[u]^{\mathcal{L}}& \mathcal{G}\ar@{.>}[u]^{\mathcal{L}}\ar@{~>}[l]^{\simeq}
}
\end{align}

Last but not least,  the quantization of the system is then represented by the irreducible unitary representations of $\mathcal{C}$ on Hilbert space, which give rise to self-adjoint operator of the algebra of classical observables. 

\subsection{Noncommutative Quantum Mechanics}
We now want to employ the quantization approach into noncommutative system, and also want to explore  the  quantization with Landau and symmetric dual gauge couplings namely dual magnetic fields. Noncommutative plane can be intepreted as such that the Lie commutator of
the positions in two-dimensional plane no longer commute. Moreover, there are several methods to achieve the noncommutative one, and it has been discussed well in \cite{gouba2016comparative}. For instance, some author introduce noncommutative plane as the positions $q^{i\prime}$ where $i=1,2$ that take over the commutative positions $q^{i}$ and momenta $p_{i}$ which known in literature as Bopp shift. While Moyal $\star$-product replacing the piecewise product.

Consider a particle on  plane $\mathbb{R}^2$ propagating through the noncommutativity system, where the phase space of the system is described by $T^\ast \mathbb{R}^2\cong \mathbb{R}^2\times \mathbb{R}^2$. The symplectic structure on $\mathbb{R}^2\times \mathbb{R}^2$ phase space is $dq^i\wedge dp_i$ and will be extended into        \begin{align}\Omega=dq^i\wedge dp_i+\frac{1}{2}\theta^{ij}dp_i\wedge dp_j,\label{SympNC}
\end{align} where $\theta^{ij}$ is anti-symmetric noncommuting parameter, $\theta^{ij}=-\theta^{ji}$ and two-dimensional plane with $i=1,2$. From the given  $\mathbb{R}^2\times \mathbb{R}^2$ phase space, we use the corresponding Hamiltonian vector field as
\begin{align}
\xi_f&= \frac{\partial f}{\partial p_i} \frac{\partial}{\partial q^i}-\frac{\partial f}{\partial q^i }\frac{\partial}{\partial p_i},\label{HVFNC}
\end{align} where the binary product of the two vector fields is Lie bracket.

This  set of vector fields (\ref{HVFNC}) using $\Omega$ generates new canonical observables
\begin{subequations}\label{canoo}
\begin{align}
\xi_{q^i}\lrcorner \Omega &= q^i-\frac{1}{2}\theta^{ij}p_j=q^{i\prime },\label{canoo1}\\
\xi_{p_i}\lrcorner \Omega &= p_i=p^{\prime}_i.
\end{align}
\end{subequations}
These observables (\ref{canoo}) proceed to develop a new Poisson bracket for noncommutative plane through the contraction of Hamiltonian vector fields $\xi_{q^i}$, $\xi_{p_i}$ and symplectic structure $\Omega$ via $\xi_{g}\lrcorner \xi_{g}\lrcorner \Omega =\{ f, g\}_{NC}$ which develops a set of the brackets for canonical observables   \begin{subequations}
	\begin{align}
\{{q^{i\prime}}, {q^{j\prime}}\} &=\theta^{ij},\\
	\{p^{\prime}_i, p^{\prime}_j\} &=0,\\
	\{ {q^{i\prime}}, p^{\prime}_j\} &=\delta^i_j,
	\end{align}\label{PoissNC} 
\end{subequations} where both new coordinates ${q^{i\prime}}$ are the extended position (\ref{canoo1}), and $p^{\prime}_i=p_i$. This set of Poisson brackets (\ref{PoissNC}) also obeys  skew-symmetric condition, Leibiniz rule and Jacobi identity. 

The symmetry of the $\mathbb{R}^2\times \mathbb{R}^2$ phase space  is Heisenberg group, $\mathtt{H}^2$ \cite{isham1984topological}. The next step is to establish the homomorphism between Heisenberg algebra  $(\text{A}^i, \text{B}_i, \text{C})\in \mathcal{L}(\mathtt{H}^2)$, and the Poisson brackets (\ref{PoissNC}) via moment map $P$ acting on the phase space. Nevertheless, the map is not homomorphism  
\begin{align} \label{nonvanishing}
\{ P^{(\text{A}^i, \text{B}_i, \text{C})}, P^{({A^{\prime}}^i, {B^{\prime}}_i, {\text{C}^{\prime}})}\}(s) -P^{[(\text{A}^i, \text{B}_i, \text{C}), ({\text{A}^{\prime}}^i, {\text{B}^{\prime}}_i, {\text{C}^{\prime}})]}(s)=\theta^{ij}(\text{B}_i\text{B}^{\prime}_j-\text{B}_j\text{B}^{\prime}_i).
\end{align}
This nonvanishing cocycle will be treated as  a central extension which immediately enlarge the algebra, and eventually Lie bracket reads
\begin{align}
[(\text{A}^i, \text{B}_i, \text{C}, \text{D}), ({\text{A}^{\prime}}^i, {\text{B}^{\prime}}_i, {\text{C}^{\prime}}, {\text{D}^{\prime}} )] =(0,0, \text{B}_i{\text{A}^{\prime}}^i-\text{A}^i{\text{B}^{\prime}}_i, \theta^{ij}(\text{B}_i\text{B}^{\prime}_j-\text{B}_j\text{B}^{\prime}_i)),\label{LieNC}
\end{align}
and the group composition                 \begin{align}
\nonumber(\text{a}^i, \text{b}_i, \text{c}, \text{d})({\text{a}^{\prime}}^i, {\text{b}^{\prime}}_i, {\text{c}^{\prime}}, {\text{d}^{\prime}})=&(\text{a}^i+{\text{a}^{\prime}}^i, \text{b}_i+{\text{b}^{\prime}}_i, \text{c}+{\text{c}^{\prime}}+\frac{1}{2}\hbar(\text{b}_i{\text{a}^{\prime}}^i  - \text{a}^i{\text{b}^{\prime}}_i),  \text{d}+{\text{d}^{\prime}}\\&\quad +\frac{1}{2} \theta ^{ij}(\text{b}_i{\text{b}^{\prime}}_j-\text{b}_j{\text{b}^{\prime}}_i) ).\label{defheis}
\end{align} As a result, the canonical group $\mathcal{C}$ of the phase space is the extended Heisenberg group $\mathtt{H}^2\rtimes \mathbb{R}$.\par 
We  will now define the operator in exponential form  acting on Hilbert space, where the Stone-von Neumann theorem still holds \cite{gouba2009uniqueness}, indicating the uniqueness of the unitary irreducible representation of the extended Heisenberg algebra is    
\begin{align}\label{unirrss}
U(a^i):= e^{-i\text{A}^i\hat{p}^\prime_i}, 
\quad
V(b_i):=e^{-i\text{B}_i{\hat{q}^{i\prime }}}, \quad
W(c,d):= e^{-i(\text{C}\hbar +\text{D}\theta)} ,                 
\end{align} where the extended position operator $\hat{q}^{i\prime} =\hat{q}^i-\frac{1}{2}\theta^{ij}\hat{p}_j$ and two central extension of the algebra are given by $\text{C}$ and $\text{D}$ with their parameter $\hbar$ and $\theta$ respectively. The operation between unitary representations (\ref{unirrss}) can be calculated with the group composition (\ref{defheis}) leading to the well-known CCR for NCQM \begin{subequations}\label{c4ccrext}
	\begin{align}
	[{\hat{q}^{i\prime }}, {\hat{q}^{j\prime }}]&= i\theta^{ij},   \\
	[\hat{p}^{\prime}_i, \hat{p}^{\prime}_j]&= 0,   \\
	[{\hat{q}^{i\prime }}, \hat{p}^{\prime}_j]&=i\hbar  \delta^i_j.
	\end{align}
\end{subequations} 

The Hilbert space of this representation is given by $L^2(\mathbb{R}^2, dq^1dq^2)$, and the representations acting on the $\psi(q^i)$ square-integrable functions of $\mathbb{R}^2$ \begin{subequations}\label{c4ncqmm}
	\begin{align}
	{\hat{q}^{i\prime }}\psi(q^i )&=\left( q^i+i\frac{1}{2}\theta^{ij}\frac{\partial}{\partial q^j} \right) \psi(q^i ),\label{c4ccrbopps}\\
	\hat{p}^{\prime}_i\psi(q^i )&=-i\hbar\frac{\partial}{\partial q^i}\psi(q^i ). 
	\end{align}
\end{subequations}

\subsection{Dual Gauge Couplings}
We now consider two choices of (dual) gauges\footnote{Dual gauge ${G}^{ij}(p_i, p_j)$ has defined as a gauge that respect to momenta instead of position for the conventional gauge} namely; Landau and symmetric  gauges of  magnetic field. The inclusion of these gauge will immediately change the group, and  its irreducible unitary representations will be computed. Dual magnetic field $ {G}$ in terms of the vector potential $ (\tilde{A}^1, \tilde{A}^2)$ is defined as
\begin{align}\nonumber 
{G}^{ij}:=& \partial^i\tilde{A}^j- \partial^j\tilde{A}^i\\
=&\frac{\partial}{\partial p_i} \tilde{A}^j   -  \frac{\partial}{\partial p_j}\tilde{A}^i.\label{c4dualgaug66}
\end{align} 
Landau dual gauge or in short, $\tilde{A}_L$, corresponding to the choice of vector potential respects to momenta 
\begin{align}\nonumber
\tilde{A}_L:=& \left(\tilde{A}^1_L, \tilde{A}^2_L\right)  
\\ =&\left(- {G} p_2,0\right). \label{cgshsj}
\end{align}
Therefore, the group generators  is modified  with the corresponding contravariant gauge;\begin{align}\label{c4ncncpos}
q^{i\prime}&=q^i+\tilde{A}^i_L(p_1, p_2).
\end{align}
From the choice of vector potential (\ref{cgshsj}) thus
\begin{subequations} 
	\begin{align}
	q^{1\prime}&=q^1- {G} p_2,\\
	q^{2\prime}&=q^2.
	\end{align}
\end{subequations}

The addition of dual magnetic field  has redefined the  position coordinate on $\mathbb{R}^2$. Consequently, the canonical group for the phase space with Landau dual gauge coupling is $\mathtt{H}^2_L$ group. The Hilbert space  is defined as follows, $\psi(q^1, q^2 )\in L^2(\mathbb{R}^2, dq^1dq^2)$, and the unitary representations of the canonical group $\mathtt{H}^2_L$ are $U(a^i)$,  $V(b_i)$ and $W(c,d)$, where those operations will be given below
\begin{subequations}  \label{eq1919}        
	\begin{align}
	U(a^1, 0)\psi(q^1, q^2 )&=\psi(q^1-a^1, q^2 ),\\
	U(0, a^2)\psi(q^1, q^2 )&=\psi(q^1, q^2-a^2 ),  \\  
	V(b_1,0)\psi(q^1, q^2 )&=e^{-iB_1q^1}\psi(q^1, q^2+ {G}b_{1} ),\\
	V(0,b_2)\psi(q^1, q^2)&= e^{-iB_2q^2}\psi(q^1, q^2),\\
	W(c, d)\psi(q^1, q^2)&=\psi(q^1, q^2)e^{i(C\hbar+D {G})}.
	\end{align}
\end{subequations}
The operations (\ref{eq1919}) correspond to                     \begin{subequations}        
	\begin{align}
	\hat{q}^{1\prime}\psi(q^1, q^2 )&= \left( q^1+ i{G}\frac{\partial}{\partial q^2} \right)\psi(q^1, q^2 )   ,\quad
	\hat{q}^{2\prime}\psi(q^1, q^2 )=  q^2 \psi(q^1, q^2 )   ,\\
	\hat{p}^\prime_1\psi(q^1, q^2 )&= -i\hbar\frac{\partial}{\partial q^1} \psi(q^1, q^2 )  ,\quad\qquad\quad
	\hat{p}^\prime_2\psi(q^1, q^2 )= -i\hbar\frac{\partial}{\partial q^2}\psi(q^1, q^2 )   .
	\end{align}
\end{subequations}           Interestingly, the CCR with Landau dual gauge appears to be similar to (\ref{c4ccrext}): 
\begin{subequations} 
	\begin{align}
	[{\hat{q}^{i\prime }}, {\hat{q}^{j\prime }}]&= i {G}^{ij},   \\
	[\hat{p}^{\prime}_i, \hat{p}^{\prime}_j]&= 0,   \\
	[{\hat{q}^{i\prime }}, \hat{p}^{\prime}_j]&=i\hbar  \delta^i_j.
	\end{align}
\end{subequations}

Symmetric dual gauge is denoted by $\tilde{A}_S$, and  chosen as 
\begin{align}\nonumber
\tilde{A}_S:=&\left(\tilde{A}^1_S, \tilde{A}^2_S\right)\\
=&\left(-\frac{ {G}}{2}p_2,\frac{ {G}}{2}p_1\right),  \label{symmm}
\end{align}
where $ {G}^{ij}=\begin{pmatrix}
0& {G}\\
- {G} &0
\end{pmatrix}$. This choice  preserves the rotational symmetry, and breaks translational symmetry in both positions $q^i$.  In terms of the group, previously it had the similar generator (\ref{unirrss}): \begin{subequations} \nonumber
	\begin{align}
	q^{1\prime}&=q^1-\frac{1}{2} {G} p_2,\\
	q^{2\prime}&=q^2+\frac{1}{2} {G} p_1.
	\end{align}
\end{subequations}

Note that the symmetric dual gauge follows the previous discussion particularly (\ref{c4ccrext}). Canonical group for the $\mathbb{R}^2\times\mathbb{R}^2$ phase space with symmetric dual gauge is $\mathtt{H}^2_S$ (where $S$ is denoted as  symmetric dual gauge). Moreover, their CCR is  also similar to (\ref{c4ccrext}), and the representation as well. Throughout the section, the result shows the equivalence between noncommutative plane, the phase space with Landau and symmetric dual gauge with regard to the canonical group      \begin{align}
\mathtt{H}^2\rtimes \mathbb{R}\cong \mathtt{H}^2_L\cong \mathtt{H}^2_S.
\end{align}

\section{Noncommutative Graphene}\label{ncg}

A free massless Dirac electron moves with Fermi velocity  $v_F\sim 10^6 ms^{-1}$, propagates through a perpendicular magnetic field $\vec{F}=F\textbf{k}$ to the xy-plane which is represented by the following Dirac equation
\begin{align}\label{dfadsg}
i\hbar	\frac{\partial \psi }{\partial t}=H_D\psi,
\end{align}

\noindent where $H_D$ is a Hamiltonian around Dirac points which are the point at the corners of Brillouin zone namely $K$ and $K^\prime$
\begin{align}\label{hamiii}
	 H_{D}=\begin{pmatrix}
	H^K& 0\\
	0 & H^{K^\prime}
	\end{pmatrix}.
\end{align}
\subsection{Magnetic Field}
Dirac-like Hamiltonian with the coupling to the vector potential $A_i$, and it is chosen by
\begin{align}\label{gauggg}
A=\left( \frac{1}{2}Fq^2, -\frac{1}{2}Fq^1\right). 
\end{align}
If we take Hamiltonian around $K$ point thus
\begin{align}
	H^K=v_F\sigma^i\cdot \mathcal{D}_i,
\end{align}
where the covariant derivative is $\mathcal{D}_i=p_i+A_i$ and Pauli matrice   \begin{align*}
	\sigma^1=\begin{pmatrix}
	0& 1\\
	1 & 0
	\end{pmatrix}, \quad\sigma^2=\begin{pmatrix}
	0& -i\\
	i & 0
	\end{pmatrix}.
\end{align*}
The Hamiltonian (\ref{hamiii}) with the coupling (\ref{gauggg}) becomes
\begin{align}\label{ffdfg}
H_D=v_F\begin{pmatrix}
0& \mathcal{D}_1-i\mathcal{D}_2 &0&0\\
\mathcal{D}_1+i\mathcal{D}_2 &  0& 0&0\\
0&0&0&\mathcal{D}_1+i\mathcal{D}_2\\
0&0&\mathcal{D}_1-i\mathcal{D}_2&0
\end{pmatrix}.
\end{align}
The eigenvalue of the wave function (\ref{dfadsg}) for Dirac $K$ point is given by
\begin{subequations}\label{ggms}
	\begin{align}
		v_F\left( \mathcal{D}_1-i\mathcal{D}_2\right)\psi^K_2&=E^K\psi^K_1,\\
		v_F\left( \mathcal{D}_1+i\mathcal{D}_2\right)\psi^K_1&=E^K\psi^K_2.
	\end{align}
\end{subequations} Note that the wave function $\psi$ corresponding to the Hamiltonian (\ref{dfadsg}) is two-component spinor $\begin{pmatrix}
\psi^K\\
\psi^{K^\prime}
\end{pmatrix}$. 
In order to calculate the energy of the components $K$ in (\ref{ggms}), thus the redefined operator requires
\begin{subequations}\label{12121}
	\begin{align}
	a&=\frac{1}{\sqrt{2\hbar F}}\left(\mathcal{D}_1+i\mathcal{D}_2 \right),\\
	a^\dagger&=\frac{1}{\sqrt{2\hbar F}}\left(\mathcal{D}_1-i\mathcal{D}_2 ,\right),
	\end{align}
\end{subequations}
which obey the following commutation relation
\begin{align}\label{ccrladder}
	\left[ a,a^\dagger \right]=1.
\end{align}
Hence the state corresponding to spinor component $K$ can be developed via creation operator as follows
\begin{align}\label{statespinor}
	\psi^K\equiv \vert n \rangle=\frac{\left(a^\dagger\right)^n}{n!}\vert 0\rangle,
\end{align}

\noindent where the ladder (\ref{12121}) and number operators respectively are
\begin{subequations}
	\begin{align}
		 a^\dagger\vert n \rangle&=\sqrt{n+1}\vert n+1 \rangle   ,\\
		 a\vert n \rangle&= \sqrt{n}   \vert n-1 \rangle   ,\\
	{	 a^\dagger a\vert n \rangle}&= n\vert n \rangle     .\label{llll}
	\end{align}
\end{subequations}
The eigenvalues of the Hamiltonian at $K$ point (\ref{dfadsg}) now can be rewritten as 
\begin{align}
	\hbar \omega \begin{pmatrix}
	0& a^\dagger\\
	a &0
	\end{pmatrix}\begin{pmatrix}
	 \psi^K_1\\\psi^K_2
	\end{pmatrix}=E^K\begin{pmatrix}
	\psi^K_1\\\psi^K_2
	\end{pmatrix},
\end{align}           where  cyclotron frequency $\omega=v_F\sqrt{\frac{2F}{\hbar}}$. Eigenvalues of the $ { a^\dagger a}$ implies      \begin{align}\label{2222}
	{a^\dagger a} \psi^K_1=a\left( \frac{E^K}{\hbar \omega}\right)\psi^K_2=\left( \frac{E^K}{\hbar \omega}\right)^2\psi^K_1.
\end{align}               Finally, comparing the number operator (\ref{llll}) with (\ref{2222}) leads to the energy spectrum of the graphene in uniform magnetic field background is           \begin{align}
	E_n=\pm \hbar v_F \frac{1}{L}\sqrt{2n},
\end{align}           where magnetic length is $L=\sqrt{\frac{\hbar}{F}}$, while positive and negative signs correspond to electrons in conduction band, and holes in valence band respectively.

\subsection{Symmetric and Landau Dual Magnetic Field}
 
The vector potential on graphene (\ref{gauggg}) now is extended with the presence of dual magnetic field (\ref{c4dualgaug66}). Therefore, the covariant derivative will be replaced by
\begin{align}\label{Pi}
	\mathcal{D}_i\rightarrow \Pi_i=p_i\left( 1+\frac{1}{2}\alpha |FG| \right) +\frac{1}{2}F_{ij}q^j,
\end{align}
where $\alpha= 0,\frac{1}{2},1$ depends on the type of gauge choices, and the magnetic field satisfies antisymmetric matrix $F_{ij}=-F_{ji}$.
Yet, the set of equations for K point (\ref{ggms}) becomes
\begin{subequations}\label{redeffoperat}
\begin{align}
v_F\left( \Pi_i-i\Pi_2\right)\psi^K_2&=E^K\psi^K_1,\\
v_F\left( \Pi_i+i\Pi_2\right)\psi^K_1&=E^K\psi^K_2.
\end{align}
\end{subequations}
Hence, in this part, the ladder operator will be further modified as        \begin{subequations}\label{newoper}
	\begin{align}
	A&=\frac{1}{\sqrt{2\hbar\gamma F}}\left(\Pi_1+i\Pi_2 \right),\\
	A^\dagger&=\frac{1}{\sqrt{2\hbar \gamma F}}\left(\Pi_1-i\Pi_2 \right),
	\end{align}
\end{subequations} where $\gamma=1+\frac{1}{2}\alpha |FG|$.
These new ladder operators respectively are creation and annihilation operators (\ref{newoper}) which also obey the commutation relation (\ref{ccrladder}). In addition, the new state corresponding to spinor component $\psi^K\equiv\vert N \rangle $  and the new number operator ${ A^\dagger A}$ can be defined in the similar way respectively as (\ref{statespinor}) and (\ref{llll}). Thus, these new defined operators will modify the Hamiltonian at $K$ point as           \begin{align}
\hbar \tilde{\omega} \begin{pmatrix}
0& A^\dagger\\
A &0
\end{pmatrix}\begin{pmatrix}
\psi^K_1\\\psi^K_2
\end{pmatrix}=E^K\begin{pmatrix}
\psi^K_1\\\psi^K_2
\end{pmatrix},
\end{align}                     where new  cyclotron frequency $\tilde{\omega}=v_F\sqrt{\left( 1+\frac{1}{2}\alpha |FG|   \right)\frac{2F}{\hbar}}=v_F\sqrt{\gamma\frac{2F}{\hbar}}$.

Symmetric dual gauge is chosen by (\ref{symmm}) which leads to the respectively corrected coefficient $\gamma$ and cyclotron frequency $\tilde{\omega}$. The eigenvalues of Hamiltonian at K point (\ref{2222}) are obtained by the definition of number operator ${ A^\dagger A}$;      \begin{align}\label{222211}
{ A^\dagger A} \psi^K_1=A\left( \frac{E^K}{\hbar \tilde{\omega}}\right)\psi^K_2=\left( \frac{E^K}{\hbar \tilde{\omega}}\right)^2\psi^K_1.
\end{align}   

In result, the spectrum energy of the graphene with symmetric dual magnetic field is
\begin{align}\label{energyE}
E_N=\pm  \hbar v_F \frac{1}{L}\sqrt{2\gamma N}.
\end{align}   

For this case, Landau dual magnetic field (\ref{cgshsj}) concerns more on operator $\Pi_1$ instead of $\Pi_2$, where the conjugate momentum (\ref{Pi}) has been modified as follows
\begin{subequations}\label{piii}
	\begin{align}
	\Pi_1&=p_1\left( 1+\frac{1}{2}\alpha |FG| \right) +\frac{1}{2}F q^2,\\
	\Pi_2&=p_2-\frac{1}{2}F q^2.
	\end{align}
\end{subequations}

However, with regard to solve the eigenvalue of the system, the ladder operator (\ref{newoper}) has been redefined using the covariant derivative (\ref{piii});\begin{align}\nonumber
A&=\frac{1}{\sqrt{2\hbar\gamma F }}\left(\Pi_1+i\Pi_2 \right),\\
A^\dagger&=\frac{1}{\sqrt{2\hbar \gamma F} }\left(\Pi_1-i\Pi_2 \right),\nonumber
\end{align}
which also obey the commutation relation (\ref{ccrladder}). However, this work follows the similar steps as what has been done in (\ref{222211}, \ref{energyE}) and it shows that the spectrum energy of the graphene with Landau dual magnetic field equivalence to the symmetric dual magnetic field one.

\section{Conclusion}\label{4aaa}
In conclusion, we have investigated the behavior of noncommutative graphene in the presence of a dual magnetic field by means of canonical group quantization. The results reveal that the canonical group for the phase space $\mathbb{R}^2\times \mathbb{R}^2$ with both symmetric and Landau dual gauge couplings are equivalent to the noncommutative plane \cite{umar2dpmssaq} namely $\mathtt{H}^2\rtimes \mathbb{R}$. Moreover, the representations of these dual gauges exhibit a similar noncommutative CCR. These findings suggest that the dual magnetic field introduces corrections to the energy spectrum of noncommutative graphene by $\pm  \hbar v_F \frac{1}{L}\sqrt{2\gamma N}$, and the use of CGQ provides a useful approach to study quantum systems in noncommutative spaces with the consideration of dual gauge couplings.

\subsection*{Acknowledgement}
The work is supported by GPUF grant, UPSI/PPPI/PYK(20200291).

\end{document}